\documentclass[epj,twocolumn]{webofc}
\usepackage[varg]{txfonts}   
\woctitle{The Innermost Regions of Relativistic Jets and Their Magnetic Fields}
\begin{document}
\title{The Parsec-Scale Structure of the Newer TeV Blazars}

\author{B. Glenn Piner\inst{1}\fnsep\thanks{\email{gpiner@whittier.edu}} \and   
        Philip G. Edwards\inst{2}}

\institute{Department of Physics and Astronomy, Whittier College, Whittier, CA, USA \and
CSIRO Astronomy and Space Science, Epping NSW, Australia}

\abstract{
We expand our previous studies of the parsec-scale structure of TeV blazars by 
presenting first-epoch images from VLBA monitoring of ten newer TeV HBLs.
All ten sources were successfully detected and imaged, and all showed a one-sided core-jet structure.
Many display a morphology common to TeV HBLs:
a short, collimated jet followed by a transition to
low surface brightness extended emission with a much broader opening angle.
The newly detected TeV HBLs tend to be fainter in the radio; the median core flux density
was 22~mJy, and the median brightness temperature was $8\times10^{9}$~K.
The brightness temperatures are well below the equipartition limit, and
thus the VLBI cores do not require strong beaming, consistent with the
modest values of Doppler and Lorentz factors found in the VLBI jets of TeV HBLs by other studies,
and contrasting with the strong beaming generally required by the TeV emission.
We study the full sample of TeV HBLs that have been observed with VLBI, and find a 
correlation between TeV flux and VLBI core brightness temperature, suggesting different
but correlated beaming factors for the TeV and radio emission. We present a discussion
of these observations in the context of velocity structures in the jets of the TeV HBLs.
}
\maketitle
\section{Introduction and Background}
\label{intro}
The number of TeV blazars has grown rapidly over the past few years;
the current generation of ground-based TeV gamma-ray telescopes has detected a total of 58 AGN 
as of this writing (tevcat.uchicago.edu). 
The majority of these (41 of 58, or about 70\%) belong to the 
class of high-frequency peaked BL Lac objects, or HBLs. 
Some of these TeV HBLs have displayed remarkable variability in their TeV gamma-ray emission on time scales 
as short as a few minutes \cite{Albert07,Aharonian07}. 
Although a variety of ideas has been proposed to explain this dramatic variability
(e.g., \cite{Begelman08,Nalewajko11,Narayan12,Barkov12}), they
share the common feature of high bulk Lorentz factors for the 
gamma-ray emitting plasma in the relativistic jets of at least $>\sim25$.
High bulk Lorentz factors are also typically used to model TeV blazar
spectral energy distributions (e.g., \cite{Tavecchio10}).

Imaging the jets of these blazars directly on the parsec-scale requires VLBI. 
Most HBLs are relatively faint in the radio, so they are not well represented in large VLBI monitoring projects. 
We have previously reported multi-epoch VLBI kinematic results for 11 established TeV HBLs
(six in \cite{Piner10}, and an additional five in \cite{Tiet12})
\footnote{These 11 sources are: Mrk 421, H 1426+428, Mrk 501, 1ES 1959+650,
PKS 2155$-$304, and 1ES 2344+514 by \cite{Piner10}, and 1ES 1101-232, Mrk 180, 1ES 1218+304,
PG 1553+113, and H 2356-309 by \cite{Tiet12}.}.

A major result of those kinematic analyses was the
absence of any rapidly moving features in the jets of TeV HBLs; all 
components in all 11 sources were either stationary or slowly moving ($<\sim1c$).
Slow apparent speeds of VLBI components in specific TeV HBLs has been confirmed
by numerous other studies (e.g., \cite{Richards13,Aleksic13,Blasi13,Lico12,Giroletti04a},
although note that TeV-detected IBLs
such as 3C~66A and BL Lac {\em do} show rapidly moving components).
While effects other than slow bulk motion can produce
slow apparent speeds of components, the complete absence of
{\em any} rapidly moving features in {\em all} of these jets, after as much as 20 years
of VLBA monitoring (for Mrk~421 and Mrk~501), and even after powerful flares
\cite{Richards13}, is very distinct from the behavior of other types of gamma-ray blazars, 
which show frequent superluminal ejections (e.g., \cite{Marscher13}).
Taken together with other measured radio properties, such as the brightness
temperatures and core dominance \cite{Lister11,Giroletti04b}, the VLBI data
imply only modest bulk Lorentz factors in the parsec-scale radio jets of TeV HBLs.
(Note that, because the sources appear one-sided on parsec scales,
the VLBI data do require that the sources be at least modestly relativistic.) 
This discrepancy between the Doppler factors estimated from the gamma-ray data and at other
wavelengths has been named the  ``Doppler Crisis'' of TeV blazars.

\begin{figure*}
\centering
\includegraphics[scale=0.95]{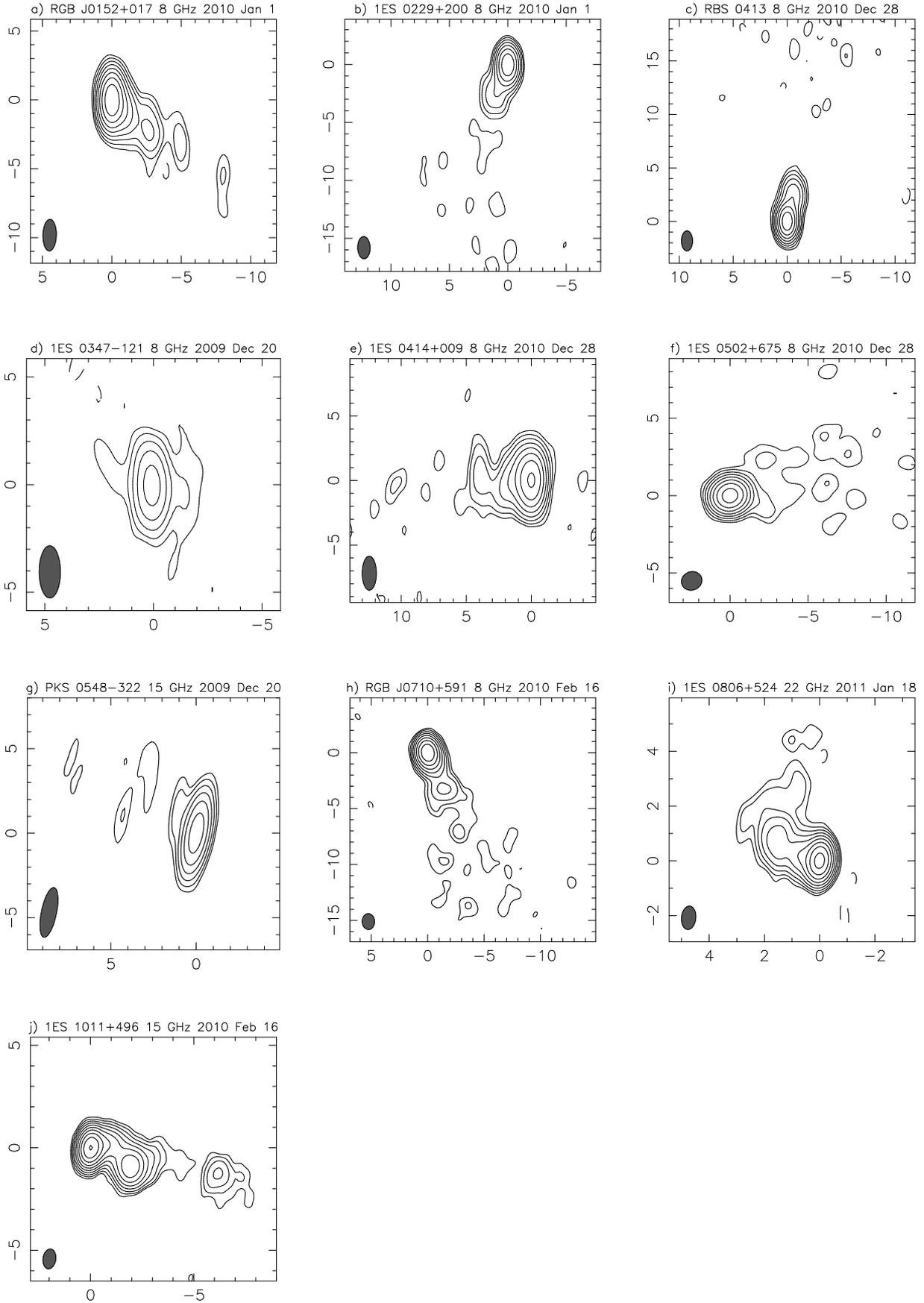}
\caption{VLBA images. Axes are in milliarcseconds.
Lowest contours are three times the rms noise level, other   
contours are each a factor of two higher.
Peak flux densities are 35, 17, 19, 5, 24, 15, 20, 28, 62, and 108 mJy beam$^{-1}$ for Figures $a$ to $j$, respectively.
Rms noise levels are 0.05, 0.05, 0.03, 0.07, 0.06, 0.03,
0.14, 0.04, 0.06, and 0.07 mJy beam$^{-1}$ for Figures $a$ to $j$, respectively.}
\end{figure*}

An obvious explanation for the Doppler Crisis is that the radio and gamma-ray emission
are produced in different parts of the jet with different bulk Lorentz factors.
Several variants of this scenario have been proposed including
decelerating jets \cite{Geor03}, spine-sheath structures \cite{Ghisellini05}, 
faster moving leading edges of blobs \cite{Lyu10},
and `minijets' within the main jet \cite{Giannios09}, 
but they all imply that the jets of HBLs contain significant velocity structures.
Some of these putative velocity structures, such as a fast spine and slow sheath, 
may under certain conditions produce observable
signatures in VLBI images, such as limb brightening of the transverse jet structure.
Limb brightening has indeed been observed 
in VLBI images of both Mrk~421 and Mrk~501 \cite{Piner09,Piner10,Giroletti04a,Giroletti06,Giroletti08}. 

These arguments for velocity structures in the jets of TeV HBLs are consistent with
recent developments in radio-loud AGN unification \cite{Meyer11,Meyer13}.
In this unification work, radio-loud AGN are divided into two
distinct sub-populations that constitute a `broken power sequence'.
The `weak' jet sub-population (corresponding to HBLs when viewed at a small angle)
follows a de-beaming curve that requires velocity gradients in the jets, such as a decelerating
or spine-sheath jet; see also the similar arguments in
earlier unification work by \cite{Chiaberge00}. The TeV HBLs may thus represent the
small viewing angle peak of a second radio-loud population with a fundamentally different
jet structure from the more powerful blazars.

We are presently
taking advantage of both the rapidly growing TeV blazar source list and the upgraded sensitivity of the 
Very Long Baseline Array (VLBA) to expand our previous work on the parsec-scale structure of TeV HBLs.
Our full sample for study now includes all HBLs in TeVCat
north of $-40^{\circ}$ (currently 38 out of 41).
In this proceeding, we present our first-epoch VLBA images of ten newer TeV HBLs discovered
during the years 2007 to 2009, several of which had never before been imaged with VLBI.

\section{Observations}
\label{obs}
We observed the ten TeV blazars RGB J0152+017, 1ES 0229+200, 
RBS 0413 (0317+185), 1ES 0347$-$121, 1ES 0414+009,
1ES 0502+675, PKS 0548$-$322, RGB J0710+591, 1ES 0806+524, and 1ES 1011+496
with the VLBA at one epoch each between 2009 and 2011, 
under observing codes BE055 and BE057. The main goal of the single-epoch
observations was to see if the sources had parsec-scale structure suitable for 
multi-epoch monitoring. Observing frequencies ranged from 8 to 22 GHz,
depending on source brightness. Earlier observations (BE055) used a data rate of 256 Mbps, 
while the later observations (BE057) used a higher data rate of 512 Mbps. 
The average observing time per epoch was five hours. 
Observations were done in phase-referencing mode for seven of the ten sources, both to aid 
in detection and to obtain precise mas-scale positions. Data were calibrated in AIPS, and imaged in Difmap.

\section{Results}
\label{results}
All ten sources were successfully detected and imaged; the VLBA images are shown in Figure~1. 
All ten sources show a one-sided core-jet structure, and we 
successfully modeled these structures with a 
circular Gaussian core and from one to four circular Gaussian jet components.
Many of the sources (most notably 1ES 0229+200, RBS 0413, 1ES 0502+675, and RGB J0710+591)
show a parsec-scale morphology that is very common among the TeV HBLs:
a collimated jet a few milliarcseconds in length, 
followed by a transition to a large region of
low surface brightness extended emission with a much broader opening angle 
beyond a few milliarcseconds from the core, suggestive of a lower-momentum flow. 
This type of morphology is reminiscent of the similar parsec-scale morphology of 
the brighter TeV blazars Mrk~421 and Mrk~501 (e.g., \cite{Giroletti06,Giroletti08}). 
At least two of the sources (1ES 0502+675 and RGB J0710+591) show a limb-brightened jet region
before the transition to low surface brightness extended emission.

\begin{figure}
\centering
\includegraphics[angle=90,scale=0.333]{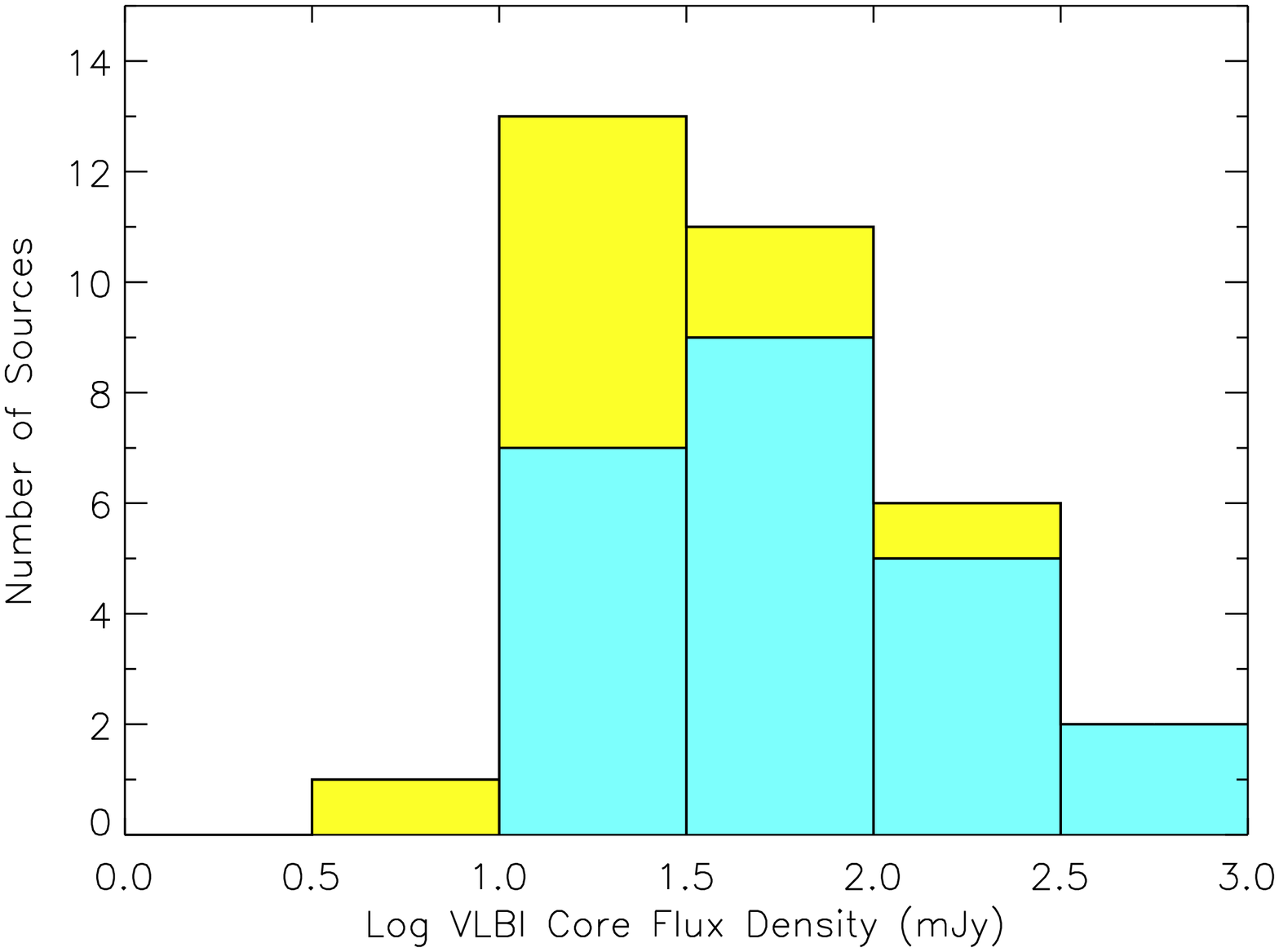}
\includegraphics[angle=90,scale=0.333]{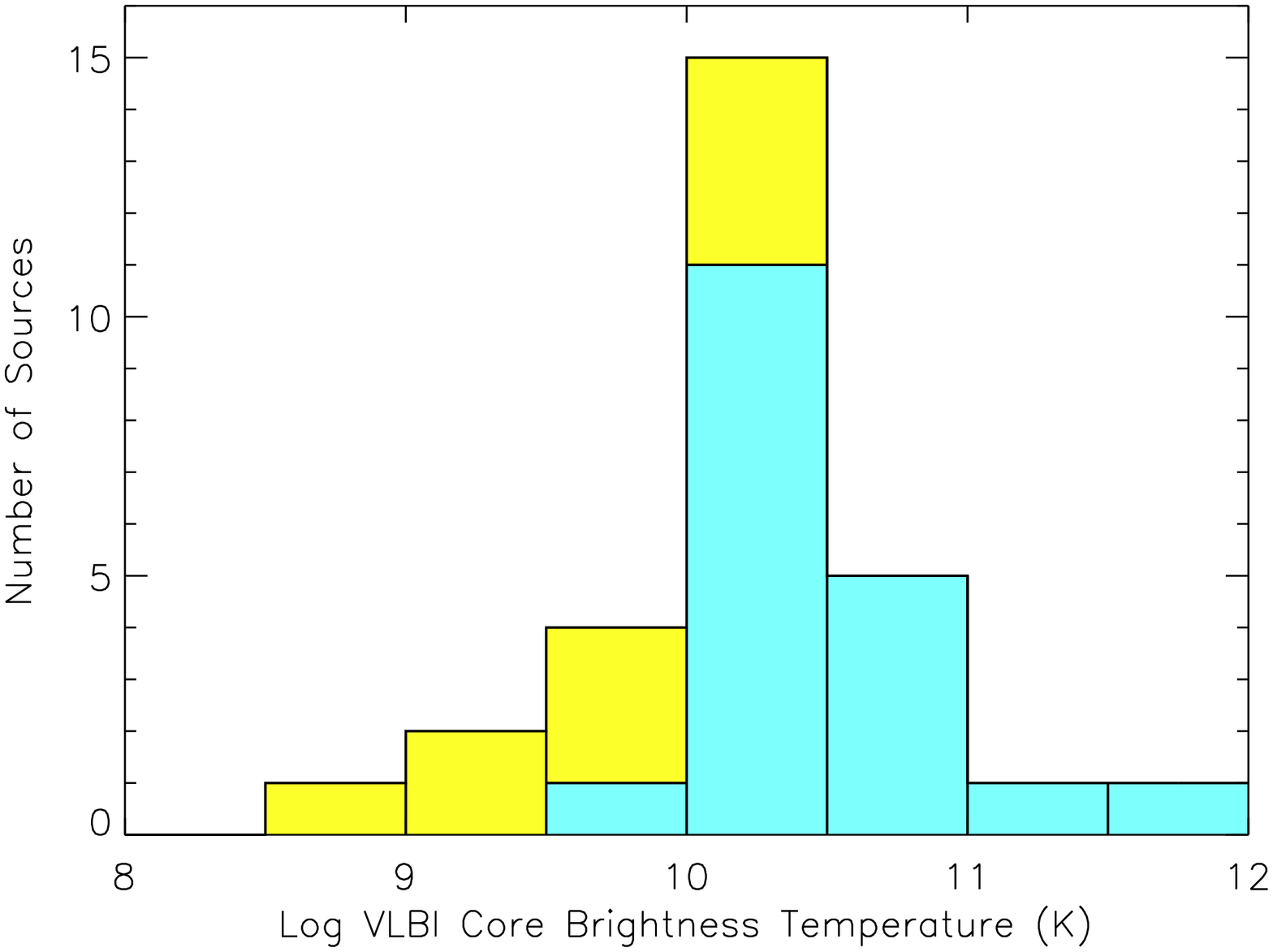}
\caption{Distributions of VLBI core flux density ($a$, top) and 
source-frame brightness temperature ($b$, bottom)
for 33 TeV HBLs. Sources from this paper are shown in yellow. Four unresolved sources do not appear
in the brightness temperature distribution.}
\end{figure}

The flux densities of the VLBI cores of these ten sources range from 
5~mJy (for 1ES 0347$-$121) to 106~mJy (for 1ES 1011+496), with a median of 22~mJy. 
These VLBI core components are partially resolved, and their source-frame brightness temperatures 
range from $8\times10^{8}$~K (for 1ES 0347$-$121) to $3\times10^{10}$~K (for 1ES 1011+496), 
with a median of $8\times10^{9}$~K
\footnote{For non-radio astronomer readers, note that brightness temperature is not a physical temperature, but a
convenient radio-astronomical means of characterizing surface brightness.}.
Figure~2 shows the distributions of core flux densities and 
source-frame brightness temperatures for all TeV HBLs imaged
with VLBI to date. The ten sources from this paper are shown in yellow.
Of the 33 sources in Figure~2$a$, the VLBI data for 21 are from our work
(\cite{Piner10,Tiet12}, and this paper), five are from the MOJAVE program,
six are from VLBI data archived at astrogeo.org, and one is from \cite{Rector03}. 
Figure~2$b$ shows the brightness temperatures 
of the 29 cores from Figure~2$a$ that are not unresolved.
For sources observed at multiple epochs, the epoch giving the median brightness
temperature has been used in Figure~2. Note in Figure~2 that
more recent TeV detections (the sources from this paper, shown in yellow) are pushing toward
fainter VLBI flux densities and lower brightness temperatures that require
greater sensitivity to successfully image.

The observed brightness temperatures of the TeV HBLs in Figure~2$b$ cluster around
relatively low values of a few times $10^{10}$~K, as was also found by \cite{Lister11}. 
They thus do not require {\em any} relativistic beaming to avoid 
either the inverse Compton or the equipartition 
brightness temperature limits \cite{Readhead94,Homan06}; however,
their one-sided morphology does imply at least modest Doppler boosting. 
As well as can be determined from the single-epoch VLBI data presented here,  
the ten new sources imaged for this paper 
display properties consistent with the relatively low values of the Lorentz
factor and Doppler factor found in the parsec-scale radio jets
of other TeV HBLs (see $\S$~\ref{intro}).

\section{Discussion}
\label{discuss}
There are at least three independent lines of evidence that velocity
structures likely exist and play a large role in the observational properties of HBLs:
\begin{enumerate}
\item{The ``Doppler Crisis'', discussed earlier~($\S$~\ref{intro}).}
\item{Unification arguments (e.g., \cite{Meyer11,Meyer13,Chiaberge00}).}
\item{Transverse features in VLBI images that may be caused by spine-sheath velocity structures
\cite{Piner09,Piner10,Giroletti04a,Giroletti06}.}
\end{enumerate} 
Given the mounting evidence for velocity structures in HBL jets, we suggest
that the so-called Doppler Crisis is actually not a crisis; the
difference in the derived values of the Doppler and Lorentz factors at different wavelengths is likely real.
Because such velocity structures are required for unification of the weak-jet sources, it would
instead be a ``crisis'' if these lower Doppler and Lorentz factors were {\em not} observable in the VLBI imaging.

\begin{figure}
\centering
\includegraphics[angle=90,scale=0.333]{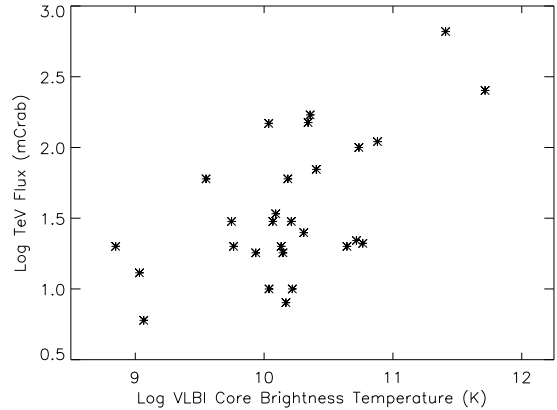}
\caption{TeV flux (in milliCrabs) versus observer-frame VLBI core
brightness temperature for 28 TeV HBLs.}
\end{figure}

If beaming factors in the TeV and radio are indeed different for the TeV HBLs,
then it is important to know whether they are different and {\em correlated} or different
and {\em uncorrelated}. In order to investigate this, 
we have computed partial correlation coefficients between the TeV flux
(taken from TeVCat for most sources; that site quotes a typical flux value for most variable sources), 
and the measured VLBI properties of the sample of TeV HBLs shown in Figure~2.
Such correlations have been reported with Fermi gamma-ray fluxes
for blazars in the MOJAVE survey (e.g., \cite{Kovalev09}), and
between the Fermi and radio fluxes for a sample or TeV blazars by \cite{Xiong13}.
We use partial Pearson correlation coefficients with the effects
of redshift removed (e.g., \cite{Padovani92}), to help avoid the effects 
of a common distance (e.g., \cite{Pavlidou12}).

We find only a weakly significant partial correlation between the TeV flux and the 
VLBI core flux density, with a significance of 0.03.
We find a much stronger partial correlation
between the TeV flux and the median VLBI core brightness
temperature, with a significance of $2.1\times10^{-3}$, 
suggesting that the {\em compactness} of the VLBI core is important.
This correlation is shown in Figure~3 for the 
28 TeV HBLs with both a TeV flux and a VLBI brightness temperature measurement.
The existence of this correlation, together with the low values for the brightness temperatures,
suggests {\em different yet correlated} Doppler factors for the VLBI core and TeV-emitting regions, 
as suggested, for example, in the model by \cite{Lyu10}.
These preliminary correlations will be investigated in more detail using the VLBA data 
for a larger sample of TeV HBLs that we are currently obtaining. 

The size scale of the VLBI core then seems to be the critical size where there are
correlations between the gamma-ray emission 
and the radio properties for the TeV HBLs. This is consistent with
observations of VLBI core variability possibly correlated with gamma-ray flares 
in Mrk~421 \cite{Richards13,Charlot06}.
However, as evidenced by the uniformly slow apparent speeds measured to date in the TeV HBLs,
such correlations seem to disappear by the scales associated with the VLBI jets.
These jets seem to be decoupled from the higher energy emission; for example,
even large flares in Mrk~421 have not typically been followed by major events in its parsec-scale jet
\cite{Richards13,Piner05}. (Note the possible exception of a superluminal {\em inward} motion 
following a large X-ray flare in Mrk~421
\cite{Niinuma12}, also see the further discussion of that result by \cite{Blasi13}.)

\section{Conclusions}
We are currently obtaining a significant amount of further VLBA data on the TeV HBLs,
which will enable us to expand the studies discussed here.
We are in the midst of multi-epoch monitoring of 
eight of the ten new sources discussed in this paper
(the other two are now included in the MOJAVE program),
and VLBI kinematic information for these eight should be available soon.

We are also using the upgraded VLBA 
to obtain both deep high-frequency images of some of the brighter TeV HBLs 
to investigate possible transverse jet velocity structures, and
first-epoch images of 12 new TeV HBLs discovered since 2010. 
This will bring the total number of TeV HBLs imaged in our program to 33.
Pending detectable parsec-scale structure in those first-epoch images, 
those 12 sources will also be followed up
with multi-epoch monitoring. All images and VLBA data files from our TeV HBL 
observing program are available online 
at:
{\footnotesize www.whittier.edu/facultypages/gpiner/research/archive/archive.html.}


\end{document}